\documentclass[conference]{IEEEtran}
\usepackage{cite}
\usepackage{graphicx}
\usepackage{amsmath,amssymb}
\usepackage{tikz}
\usetikzlibrary{shapes.geometric,arrows.meta,positioning,fit,backgrounds}
\usepackage{booktabs}
\usepackage{multirow}
\usepackage{xcolor}
\PassOptionsToPackage{hyphens}{url}
\usepackage[hidelinks]{hyperref}
\begin{document}

\title{Closing the Loop: An Access-Control Architecture for Automated, Anomaly-Driven Network Revocation in IoT Deployments}

\author{
\IEEEauthorblockN{Muhammet Emir Korkmaz\IEEEauthorrefmark{1}, Kemal Bicakci\IEEEauthorrefmark{1}\IEEEauthorrefmark{2}, Yusuf Uzunay\IEEEauthorrefmark{1}}
\IEEEauthorblockA{\IEEEauthorrefmark{1}Securify Information Technology and Security Training Consulting Inc., 06378, Ankara, T\"URK\.IYE}
\IEEEauthorblockA{\IEEEauthorrefmark{2}Informatics Institute, Istanbul Technical University, Ayazaga Campus, 34467, Maslak, Istanbul, T\"URK\.IYE}
\IEEEauthorblockA{Corresponding author: kemal.bicakci@securifyidentity.com}
}

\maketitle

\begin{abstract}
Network-based anomaly detection for IoT devices has matured to the point of reporting strong detection accuracy, yet most published systems stop at raising an alert and leave the question of automated enforcement to future work or to a programmable data plane that few real networks operate. This paper presents an access-control architecture that closes that loop using only standard, already-deployed protocols. Devices authenticate via IEEE 802.1X with EAP-TLS, and a RADIUS server acts as a continuous policy decision point capable of evicting an active session via a Change-of-Authorization Disconnect-Request and permanently excluding a device through certificate revocation. A central, contextual access policy engine continuously consumes the anomaly detector's output and actuates this response over a narrowly restricted channel to the RADIUS server; the same engine is designed to be extensible to other access types, though this paper evaluates only the network access-control mechanism. This mechanism is driven by an anomaly signal from a one-class detector adapted from a prior MUD/SDN-based design, replacing its per-flow multi-model pipeline with passive traffic capture and a single fused model that combines a cluster-based, a volumetric, and a protocol-signature score. On a single testbed device, the detector reaches an AUC of 0.9964 and detects all 24 evaluated attack scenarios (eight attack types at three intensities) using roughly 43$\times$ less training data than the reference design, and the resulting alerts reliably trigger the automated disconnect-then-revoke response, which we measure to evict a device from the network in 335.8\,ms on average and complete certificate revocation in a further 111.5\,ms. We report this evaluation as a demonstration of the closed-loop architecture rather than of the detector itself, and discuss multi-device generalization as a concrete next step.
\end{abstract}


\section{Introduction}
\label{sec:introduction}

The number and diversity of Internet of Things (IoT) devices connected to enterprise and industrial networks has grown far faster than the security tooling available to manage them. Security cameras, smart sensors, printers, and building-automation equipment now sit on the same networks as traditional IT assets, yet the vast majority of these devices offer no host-based defenses, receive infrequent firmware updates, and cannot run conventional endpoint security agents. The consequences of this imbalance are well documented: botnets such as Mirai compromised hundreds of thousands of consumer and enterprise IoT devices to launch some of the largest distributed denial-of-service attacks on record~\cite{antonakakis2017mirai}, and volumetric attacks that abuse or originate from IoT devices remain a persistent threat to both the devices themselves and the networks that host them.

The research community has responded with a large body of work on network-based anomaly detection for IoT traffic. Because IoT devices exhibit a narrow, repetitive set of behaviors compared to general-purpose computers, one-class and unsupervised learning methods trained solely on benign traffic have proven effective at flagging deviations that correspond to attacks~\cite{hamza2019detecting,meidan2018nbaiot,mirsky2018kitsune}. A parallel line of work has used the IETF Manufacturer Usage Description (MUD) standard~\cite{lear2019rfc8520} together with Software-Defined Networking (SDN) to formally constrain and monitor the traffic a device is allowed to generate~\cite{hamza2018combining,hamza2022verifying,singh2019clearer,feraudo2020sok}.

What is comparatively rare in this literature is the other half of the problem: once an anomaly has been detected, what actually happens to the device? Most published systems stop at producing an alert or a classification label, evaluated against metrics such as precision, recall, or the area under the ROC curve. Turning a detection signal into an enforceable network action -- disconnecting the device, restricting what it can reach, or permanently excluding it -- is typically left as future work, or is assumed to require a programmable data plane (SDN/OpenFlow) and a MUD profile that, in practice, the large majority of commercial IoT device manufacturers do not publish~\cite{feraudo2020sok}.

This paper presents a closed-loop architecture that treats anomaly detection as one input to an access-control decision rather than as the end product. IoT devices authenticate onto the network via IEEE 802.1X~\cite{ieee8021x2020} with certificate-based EAP-TLS, with a RADIUS server acting as the network policy decision point. Critically, RADIUS's dynamic authorization extensions~\cite{chiba2008rfc5176} let this membership be acted on after the fact, not only at connection time: a central access policy engine continuously consumes the anomaly detector's output and, when an alert fires, sends a Change-of-Authorization/Disconnect-Request over a narrowly restricted channel to evict the device from the network, while certificate revocation through a periodically refreshed revocation list provides a permanent block that survives reauthentication attempts. The same access policy engine is designed to be extensible to other access types under the same contextual rule model, though we evaluate only the network access-control mechanism in this paper.

The anomaly detector that feeds this architecture is deliberately lightweight rather than novel: it adapts the one-class clustering approach of Hamza et al.~\cite{hamza2019detecting}, replacing their MUD/SDN-derived flow telemetry with a controlled benign-traffic generator and passive packet capture, and replacing their per-flow ensemble of models with a single combined model that fuses a cluster-based score, a volumetric score, and a protocol-signature score. We report this design and its evaluation not as the paper's central claim, but because it is the trigger that makes the architecture's response path observable end to end: eight attack types at three intensity levels are detected and, in the tested network-level path, met with an automated disconnect-then-revoke response.

The contributions of this paper are as follows:
\begin{itemize}
\item An access-control architecture for IoT that closes the loop between network-behavior anomaly detection and automated enforcement, built entirely on standard, widely deployed protocols (802.1X/EAP-TLS/RADIUS) without requiring SDN/OpenFlow infrastructure or manufacturer-published MUD profiles, and designed around a central policy engine extensible to other access types.
\item An end-to-end demonstration of the closed loop on a physical testbed: detection of an anomaly triggers an automated, two-stage response -- an immediate RADIUS CoA Disconnect-Request followed by permanent exclusion via certificate revocation -- evaluated across eight attack types at three traffic intensities.
\item A data-efficient, unified instantiation of one-class IoT anomaly detection that replaces a MUD/SDN-derived multi-model pipeline with passive capture and a single fused model, reaching comparable detection performance (AUC~$=0.9964$) with roughly 43$\times$ less benign training data than the architecture it adapts.
\item A contextual policy model, expressed as a small set of condition-action rules, that generalizes beyond binary alarms to a range of dynamic actions (deny, restrict) and that unifies audit logging across the architecture.
\end{itemize}

The remainder of the paper is organized as follows. Section~\ref{sec:related-work} reviews related work on MUD/SDN-based enforcement, dynamic network authorization, and one-class IoT anomaly detection. Section~\ref{sec:architecture} describes the access-control architecture and its policy model. Section~\ref{sec:anomaly-detection} details the anomaly detector that triggers the architecture. Section~\ref{sec:response-pipeline} describes the automated response pipeline. Section~\ref{sec:experimental-setup} and Section~\ref{sec:results} present the experimental setup and results. Section~\ref{sec:discussion} discusses the design relative to the reference work, Section~\ref{sec:limitations} states the limitations of the current evaluation, and Section~\ref{sec:conclusion} concludes.

\section{Related Work}
\label{sec:related-work}

\subsection{MUD- and SDN-Based Enforcement}
The IETF Manufacturer Usage Description (MUD) standard~\cite{lear2019rfc8520} lets a device manufacturer publish a formal description of the network behavior a device is expected to exhibit, which the network can then translate into access-control rules. Hamza et al.\ were among the first to combine MUD with SDN, translating MUD policies into OpenFlow rules and forwarding only the resulting exception traffic to an intrusion detection system~\cite{hamza2018combining}, and later extending this into a full volumetric-attack detector that trains one-class models on coarse- and fine-grained SDN telemetry gathered per MUD-compliant flow~\cite{hamza2019detecting}. A follow-up work formalizes MUD profile verification and monitoring, checking generated profiles for internal consistency and compatibility with organizational policy~\cite{hamza2022verifying}. Singh et al.\ extend the MUD model to devices that do not natively support it by learning a behavioral profile directly from captured traffic~\cite{singh2019clearer}. A 2020 survey highlights the practical barriers to MUD adoption, chief among them the fact that most commercial IoT manufacturers do not publish MUD files for their products, forcing adopters either to forgo MUD-based enforcement or to derive profiles themselves~\cite{feraudo2020sok}. This gap directly motivates our decision to forgo MUD/SDN dependence and instead generate known-good protocol traffic under our own control.

Beyond MUD, SDN has also been used more generally to isolate or quarantine misbehaving IoT devices. Candal-Ventureira et al.\ propose an NFV/SDN architecture that routes suspicious devices into a dedicated quarantine network slice for deeper inspection once a lightweight first-stage detector raises a flag~\cite{candal2020quarantining}. Like the MUD-based systems above, this line of work assumes a programmable data plane is available end to end, which is a reasonable assumption in an operator-controlled 5G core but a stronger one in a typical enterprise or industrial LAN built from off-the-shelf switches and access points.

\subsection{Dynamic Network Authorization}
Our response pipeline builds on two long-standing IETF mechanisms rather than a programmable data plane. IEEE 802.1X~\cite{ieee8021x2020} defines port-based network access control in terms of a supplicant, an authenticator, and an authentication server, and is already widely deployed for enterprise Wi-Fi and wired access. RADIUS's Change-of-Authorization and Disconnect-Request extensions~\cite{chiba2008rfc5176} allow an authorization server to modify or terminate an already-authenticated session without waiting for the client to reauthenticate, a capability that is normally used for accounting or policy-refresh purposes but that we repurpose as the network-eviction primitive triggered by an anomaly signal. Because 802.1X and RADIUS CoA are standard features of commodity access points and RADIUS servers, the resulting enforcement path does not require an OpenFlow-capable switch, an SDN controller, or a MUD profile -- only a RADIUS server willing to issue a CoA request when instructed to.

\subsection{One-Class and Unsupervised IoT Anomaly Detection}
A separate thread of work focuses purely on the detection problem, independent of any enforcement mechanism. Meidan et al.'s N-BaIoT trains deep autoencoders on benign traffic snapshots from commercial IoT devices and flags anomalies via reconstruction error, demonstrating strong performance against Mirai and BASHLITE infections~\cite{meidan2018nbaiot}. Mirsky et al.'s Kitsune uses an ensemble of lightweight autoencoders to perform fully unsupervised, online intrusion detection on resource-constrained gateways~\cite{mirsky2018kitsune}. These systems, along with the MUD/SDN-based detector of Hamza et al.~\cite{hamza2019detecting} that we adapt directly (Section~\ref{sec:anomaly-detection}), share the same one-class philosophy we adopt -- training exclusively on benign traffic and treating deviation from the learned normal region as the anomaly signal -- but, like the MUD/SDN literature above, none of them close the loop with an automated, evaluated enforcement action.

\subsection{Commercial Identity Threat Detection and Response}
\label{subsec:itdr}
Outside the academic literature, RADIUS-based automated enforcement is already an active commercial category. Gartner named Identity Threat Detection and Response (ITDR) a distinct security discipline in 2022, covering tools that detect compromise of identity systems and respond once preventive access controls have been bypassed~\cite{gartner2022itdr}. Cloud RADIUS vendors have begun applying this concept specifically to wireless and IoT network access: IronWiFi's WiFi ITDR product, for example, analyzes RADIUS authentication and accounting events to build per-identity behavioral baselines and, on detecting an anomaly, triggers an automated response through RADIUS Change-of-Authorization, session blocking, or certificate revocation~\cite{ironwifi2026itdr} -- the same enforcement primitives used in this paper. This is a useful data point: it confirms that RADIUS CoA and certificate revocation are viewed industry-wide as practical, deployable enforcement mechanisms, not merely an academic convenience. However, these products detect anomalies in \emph{identity and session behavior} -- unusual login times, atypical locations, device-fingerprint changes, credential misuse -- derived from authentication and accounting events at the RADIUS server. They do not analyze the network traffic a device sends or receives after it has authenticated. A device with a valid, unrevoked credential that is subsequently compromised and repurposed for a volumetric or reflection attack (Section~\ref{subsec:attack-scenarios}) would not appear anomalous to an identity-behavior baseline, since its authentication pattern is unchanged; catching it requires visibility into the traffic itself, which is the gap our passively captured, one-class traffic detector is built to fill.

\subsection{Positioning of This Work}
The systems reviewed above address enforcement infrastructure that presumes a programmable network, detection accuracy in isolation, or identity-behavior anomalies rather than network-traffic behavior. The architecture presented in this paper reports a measured path from a network-\emph{traffic}-behavior anomaly to a concrete access-control action -- session termination and certificate revocation -- built entirely on protocols (802.1X, RADIUS CoA) that are already present in most enterprise networks and, as Section~\ref{subsec:itdr} shows, are already used by industry for identity-anomaly-driven enforcement; our contribution is to drive the same enforcement primitives from network-traffic behavior instead, which catches a different and complementary class of attack. The anomaly detector we use to drive this pipeline is intentionally derived from prior work~\cite{hamza2019detecting} rather than novel; our contribution lies in the architecture that consumes its output and in demonstrating that a single, unified one-class model can replace a MUD/SDN-derived multi-model pipeline at a fraction of the training-data cost (Section~\ref{sec:discussion}).

\section{System Architecture}
\label{sec:architecture}

\begin{figure*}[t]
\centering
\includegraphics[width=0.75\linewidth]{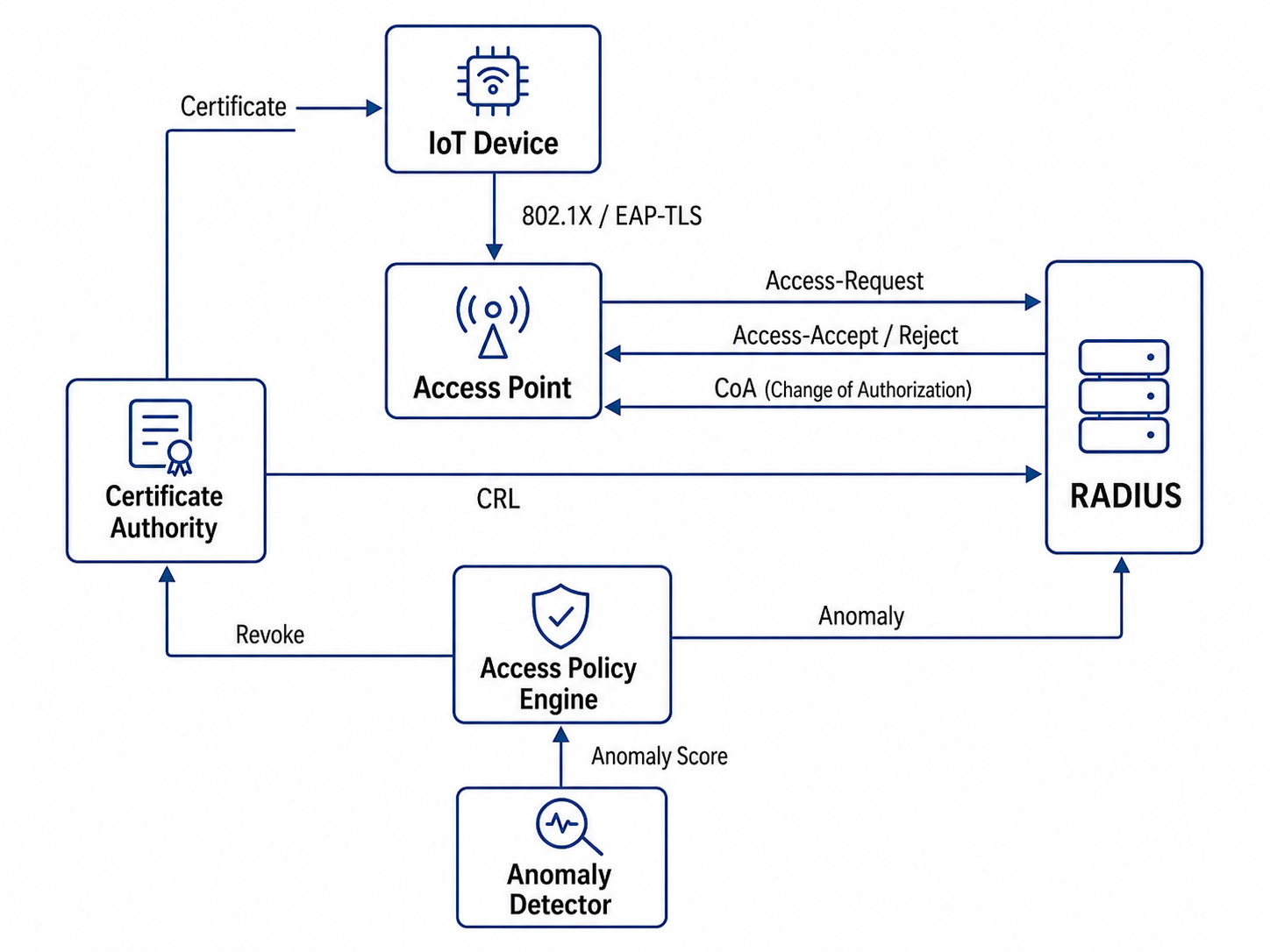}
\caption{System architecture. The IoT device presents a certificate issued by the certificate authority and authenticates via 802.1X/EAP-TLS through the access point; RADIUS grants or rejects access and separately receives the certificate revocation list (CRL) from the certificate authority. The anomaly detector's score is consumed by the central access policy engine, which, on the same alert, signals RADIUS to issue a Change-of-Authorization (CoA) request against an active session and instructs the certificate authority to revoke the device's certificate -- both actions triggered automatically and concurrently, with no separate confirmation step.}
\label{fig:architecture-overview}
\end{figure*}

The architecture is designed around a single principle: access decisions for IoT devices should be evaluated centrally and continuously, not only at the moment a device joins the network. Figure~\ref{fig:architecture-overview} summarizes the components involved. A device's network membership is governed by standard AAA infrastructure -- an access point, a RADIUS server, and a certificate authority -- while a central access policy engine continuously consumes the anomaly detector's output and, when warranted, acts on that membership through the same infrastructure. This is the mechanism evaluated experimentally in Sections~\ref{sec:experimental-setup}--\ref{sec:results}. The access policy engine is designed to also govern other kinds of access under the same contextual model; those other access types are outside the scope of the evaluation reported here.

\subsection{Network Access Control}
\label{subsec:network-level}
Network membership follows the standard IEEE 802.1X model~\cite{ieee8021x2020}: the IoT device is the supplicant, the wireless or wired access point is the authenticator, and a RADIUS server acts as the authentication server and, in our design, as the network policy decision point. Devices authenticate via EAP-TLS, each provisioned with a unique client certificate issued by a lab certificate authority. The certificate authority also publishes the certificate revocation list (CRL) that RADIUS consults both at authentication time and, as described below, as a durable enforcement mechanism.

Access control at this level is not confined to the initial authentication handshake. RADIUS's dynamic authorization extensions~\cite{chiba2008rfc5176} allow an already-authenticated session to be acted on later: a Change-of-Authorization (CoA) request can tighten the policy applied to a session, and a Disconnect-Request can terminate it outright. The next subsection describes how these dynamic requests are actually triggered in our implementation.

\subsection{Anomaly-Driven Response}
\label{subsec:response-channel}
When the alarm state machine (Section~\ref{subsec:live-monitoring}) signals alert-onset, its output is consumed by the access policy engine, which actuates a response over a deliberately narrow channel: the engine connects to the RADIUS server over a pre-configured SSH channel, restricted server-side by a wrapper script to a single operation -- issuing a CoA Disconnect-Request for a specified MAC address (Section~\ref{subsec:instant-disconnect}). This channel exists solely to carry out policy rule P-03 (Table~\ref{tab:network-policies}); it carries no other traffic and grants no shell access beyond that one operation. The same alert-onset transition also triggers, concurrently and without a separate confirmation step, a second action over a different channel: the engine calls a token-authenticated HTTP endpoint on the certificate authority (validated via an \texttt{X-CA-TOKEN} header) to request revocation of the device's certificate (Section~\ref{subsec:cert-revocation}), which the certificate authority then reflects in the CRL that RADIUS consults. The two channels are each independently access-controlled -- the wrapper script gates the disconnect channel, and the token-authenticated endpoint gates the revocation channel -- and together form the concrete mechanism through which the anomaly detector's output becomes a network action.

\subsection{Contextual Policy Model}
\label{subsec:policy-model}
Access decisions are governed by a small set of condition-action rules evaluated against contextual attributes of the request, rather than a static allow-list: the authentication method used, the network segment or source address the request originates from, the current state of the requesting device, and the time of the request. Representative rules are shown in Table~\ref{tab:network-policies}. Of these, P-01, P-02, and P-04 reflect RADIUS's and EAP-TLS's own native decision logic -- certificate validity, CRL membership, and network-segment checks already carried out as part of standard authentication -- and required no additional enforcement logic on our part. The one rule for which this paper contributes custom-built, tested enforcement logic is P-03, whose anomaly-triggered action chain (Section~\ref{sec:response-pipeline}) is actuated automatically whenever the anomaly detector raises an alert. The same access policy engine that hosts this rule set is designed to be extensible to other access types governed by an analogous contextual model, though evaluating those is outside the scope of this paper.

\begin{table}[t]
\centering
\caption{Representative network-level access policies.}
\label{tab:network-policies}
\small
\begin{tabular}{@{}p{0.11\linewidth}p{0.45\linewidth}p{0.34\linewidth}@{}}
\toprule
\textbf{ID} & \textbf{Condition} & \textbf{Action} \\
\midrule
P-01 & Successful EAP-TLS authentication & Grant network access \\
P-02 & Access request from an unregistered device & Deny access \\
P-03 & Anomaly detected in network behavior & Restrict or terminate the active session \\
P-04 & Access request from an unauthorized network segment & Deny access \\
\bottomrule
\end{tabular}
\end{table}

\subsection{Audit and Traceability}
\label{subsec:audit}
Every decision is logged to a unified audit trail rather than being scattered across per-component logs. RADIUS records capture Access-Request/Access-Accept/Access-Reject exchanges and accounting data; response records capture every invocation of the restricted SSH channel together with the CoA/Disconnect-Request it triggered. Consolidating these sources lets an analyst reconstruct, for any device, a single timeline spanning its network join events and any anomaly-driven interventions -- which is also what makes the response-latency measurements in Section~\ref{subsec:response-latency} possible to extract after the fact.

\subsection{Design Rationale: Why Not SDN?}
\label{subsec:why-not-sdn}
An alternative design, followed by much of the MUD-based literature (Section~\ref{sec:related-work}), would enforce access decisions by installing flow rules on an OpenFlow-capable switch under an SDN controller. We deliberately avoid this dependency for two reasons relevant to real deployments. First, it presumes a programmable data plane throughout the network, which is not the case for the commodity access points and switches found in most enterprise and industrial LANs. Second, SDN-based enforcement of MUD policies presumes the availability of a MUD profile for each device, and, as noted in Section~\ref{sec:related-work}, most commercial IoT manufacturers do not publish one. Building this architecture on 802.1X and RADIUS CoA instead means it can be deployed on top of AAA infrastructure that many organizations already operate, at the cost of coarser-grained enforcement (session-level eviction rather than per-flow rule installation).

\section{Anomaly Detection Component}
\label{sec:anomaly-detection}

\begin{figure}[t]
\centering
\includegraphics[width=0.95\linewidth]{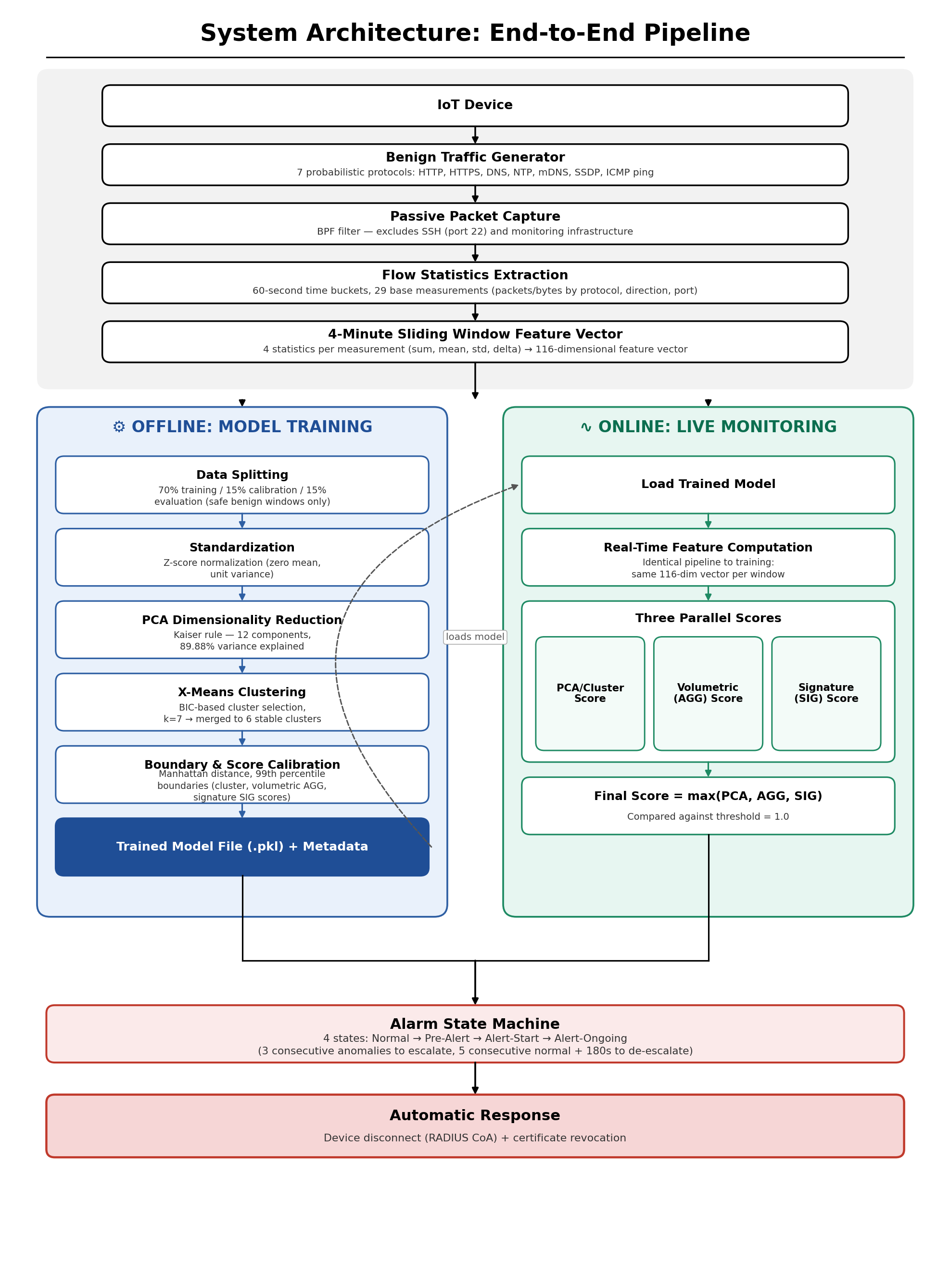}
\caption{Overall data flow of the anomaly-detection pipeline, from controlled benign-traffic generation through passive capture, flow-statistics extraction, and one-class model training/calibration to live scoring.}
\label{fig:sys-architecture}
\end{figure}

The anomaly detector is the component that generates the trigger consumed by the access-control architecture of Section~\ref{sec:architecture}. It follows the one-class classification principle of Hamza et al.~\cite{hamza2019detecting}: the model is trained exclusively on benign traffic, and an anomaly is defined as a deviation from the learned boundary of normal behavior, without requiring labeled attack samples during training. Figure~\ref{fig:sys-architecture} shows the overall data flow, from benign traffic generation through feature extraction to model training and live scoring.

\subsection{Controlled Benign-Traffic Generation}
\label{subsec:benign-traffic}
Unlike the reference architecture, which derives per-flow telemetry from SDN switches operating under a MUD profile, our design has no SDN data plane to draw traffic statistics from (Section~\ref{subsec:why-not-sdn}). Instead, a benign-traffic generator runs on the monitored device itself and produces real, protocol-level requests across seven categories that are representative of typical IoT network behavior: HTTP, HTTPS, DNS, NTP, mDNS, SSDP, and ICMP echo (ping). This traffic is captured passively at the network interface through a BPF filter that admits only packets to or from the monitored device's IP address, while excluding SSH management traffic and traffic from the monitoring infrastructure itself, so that the recorded trace reflects only the device's own network activity.

The choice of these seven categories follows from how consumer and industrial IoT devices are known to behave on a network: periodic DNS queries to a local resolver, NTP-based time synchronization, mDNS/SSDP-based local service discovery, connectivity-check pings, and HTTP/HTTPS requests to cloud services. This is the same behavioral profile that a Manufacturer Usage Description would formally encode, but because MUD profiles are rarely published (Section~\ref{sec:related-work}), we generate matching traffic directly rather than deriving it from a profile.

Each of the seven protocols contributes to a different region of the feature space described in Section~\ref{subsec:features}. HTTP/HTTPS traffic shapes the benign distribution of aggregate local and Internet-facing packet/byte counters; DNS and NTP queries, both carried over UDP, jointly establish the benign band for general inbound and outbound UDP counters that a UDP-flood attack would later push outside; mDNS traffic (multicast PTR queries to 224.0.0.251) also contributes to these general UDP counters, while SSDP traffic (M-SEARCH messages to 239.255.255.250:1900) is tracked through its own dedicated port counter and is the primary source of the benign baseline that an SSDP reflection attack would violate; and ICMP echo requests to both local and Internet addresses establish the benign band for inbound/outbound ICMP counters relevant to Ping-of-Death and Smurf-style attacks.

A second design element is stochastic execution. Hamza et al.\ report that a model trained on only two days of benign data achieved under 50\% accuracy, rising above 96\% once an additional day contributing new benign states was included~\cite{hamza2019detecting}, underscoring that detector quality depends on the diversity of benign states seen during training rather than on data volume alone. Our generator addresses this by probabilistically deciding, on each cycle, which protocols run (HTTP with probability 0.9, NTP with 0.5, mDNS and SSDP with 0.4 each), how many requests each active protocol issues, and how long to wait between requests -- so that consecutive one-minute windows differ in request density and protocol mix even though they draw from the same seven-category vocabulary. This prevents the model from collapsing the normal region onto an unrealistically narrow pattern and from flagging benign timing jitter as anomalous. The generator also aligns its execution to Unix-epoch minute boundaries so that its output lines up exactly with the fixed-size windows used by the flow-statistics extractor (Section~\ref{subsec:features}).

\subsection{Data Collection and Labeling}
\label{subsec:data-collection}
Because the model is trained exclusively on benign data, it is essential that no attack-period traffic contaminate the training set, while a separate, reliable record of attack timing is still needed to measure detection performance after the fact. A labeling component runs alongside packet capture and records the start and end Unix timestamps of every attack phase to a label file, updated incrementally after each phase so that a partially completed experiment retains valid labels for whatever phases did complete.

Data collection follows a structured protocol of alternating benign, attack, and transition phases: a long opening benign phase establishes the baseline traffic distribution, followed by a sequence of attack phases -- each run at low, medium, and high intensity (1, 10, and 100 packets/second, following the same three-tier severity convention used in the reference work) -- with an intervening transition phase after every attack to prevent residual effects from bleeding into the next phase, and a long closing benign phase. After capture, packets observed within a 5\,ms window of one another and matching on source/destination address, protocol, port, and size are de-duplicated to remove artifacts of duplicate link-layer capture before the trace is handed to the feature-extraction stage.

\subsection{Flow-Statistics Extraction and Feature Construction}
\label{subsec:features}
The cleaned packet trace is converted into fixed-size time windows of 60 seconds, using the same modular-arithmetic timestamp binning in both the offline extractor and the live monitor (Section~\ref{subsec:live-monitoring}) so that window boundaries are always reproducible from a raw timestamp. Every packet that passes the same filtering rules used during capture is classified along two axes: direction (device-to-local, local-to-device, device-to-Internet, Internet-to-device) and protocol/port (TCP, UDP, ICMP, and ARP, with TCP/UDP further keyed by a reference set of well-known service ports covering DNS, NTP, HTTP, HTTPS, SNMP, and SSDP; ephemeral ports are deliberately excluded to avoid injecting high-variance, low-information noise into the feature space). TCP SYN packets without the ACK flag are additionally tracked in a dedicated counter, separate from general TCP counters, because inbound SYN volume is a strong, low-noise signature of SYN-flood and reflection attacks, whereas outbound SYN activity is treated as weak evidence only, since it also occurs during normal IoT cloud-connectivity attempts. Flow counts per window are derived using a flow-timeout heuristic, and windows with no observed traffic are still emitted as zero-valued rows so that the resulting time series has no gaps that would break label alignment. This produces a 29-dimensional base measurement per window, spanning packet, byte, and flow counters across the direction/protocol categories above.

\begin{figure}[t]
\centering
\includegraphics[width=0.85\linewidth]{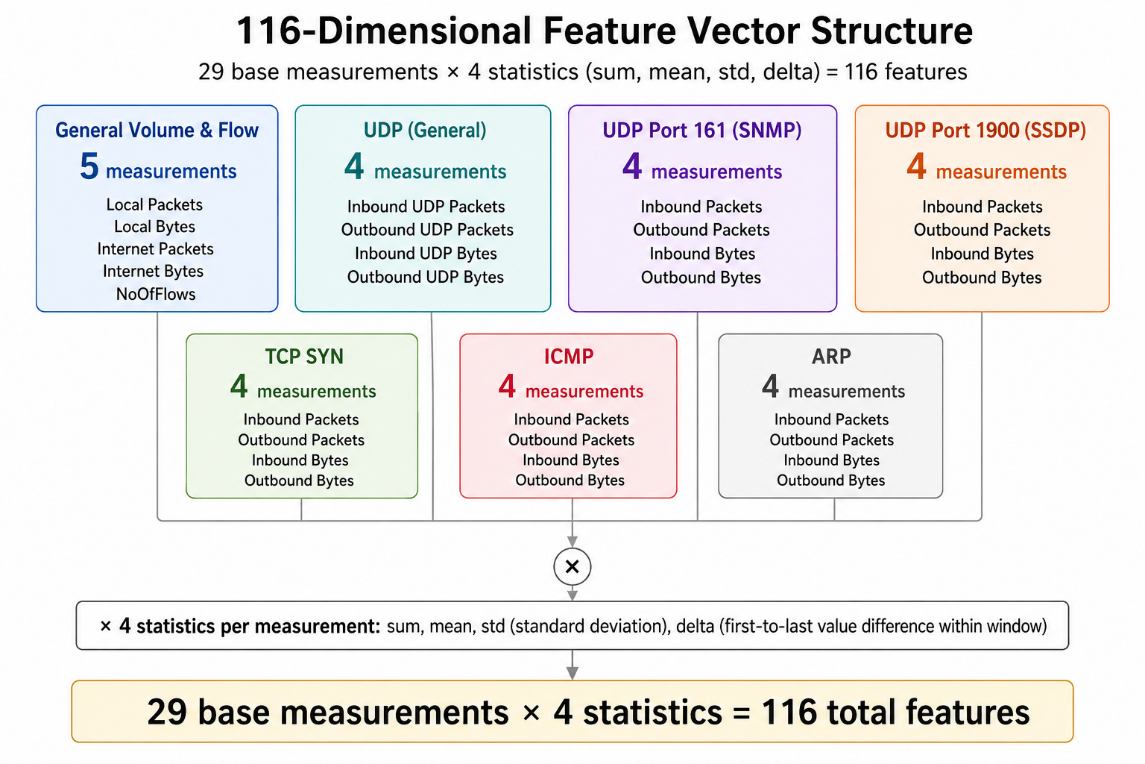}
\caption{Structure of the 116-dimensional feature vector: four rolling-window statistics (sum, mean, standard deviation, delta) computed over each of the 29 base measurements.}
\label{fig:feature-vector}
\end{figure}

These per-window base measurements are not fed to the model directly. A sliding window of four consecutive one-minute time bins is used to compute, for each of the 29 base measurements, four statistics: the sum, mean, standard deviation, and the delta between the first and last value in the window (Figure~\ref{fig:feature-vector}). Following the reference work's evaluation of window sizes, a four-minute window was adopted as it accumulates enough signal to detect low-rate attacks while keeping state-memory requirements modest; larger windows were found not to justify their added cost~\cite{hamza2019detecting}. The sum/mean/standard-deviation triplet follows the reference design directly, and the delta statistic is an addition intended to make short-term trend changes within the window directly visible to the model. Because attack and transition windows are excluded from the safe benign set (Section~\ref{subsec:training}), the remaining benign windows are not a single continuous time series but a set of disjoint blocks separated at each excluded region; the windowing procedure detects these discontinuities and never slides a window across a block boundary, which would otherwise mix unrelated periods into a single, meaningless statistic. The result is a 116-dimensional feature vector ($29 \times 4$) per admissible window.

\subsection{Detector Training and Calibration}
\label{subsec:training}
Every window is first checked against the label file: windows overlapping an attack interval, or falling within a configurable buffer of transition windows immediately before and after an attack, are excluded from training, since traffic in these transition regions is neither reliably benign nor reliably representative of the attack. The remaining, safe benign windows are split chronologically into a 70\% training set and two 15\% holdout sets used for boundary calibration and evaluation, respectively.

Because the 116 raw features span very different numeric ranges, each is standardized to zero mean and unit variance using parameters estimated only from the training split; the calibration and evaluation splits are transformed with these parameters but never used to estimate them, which prevents information from otherwise "unseen" data from leaking into the scaling step and inflating apparent performance. Principal component analysis is then applied to the standardized training features, retaining components with eigenvalue greater than one under the Kaiser criterion (with a minimum of two components enforced), following the reference work's own use of this criterion~\cite{hamza2019detecting}; in our deployment, the Kaiser criterion selected 12 components, jointly explaining 89.88\% of the variance in the original 116-dimensional space.

\begin{table}[t]
\centering
\caption{Bayesian Information Criterion (BIC) score by candidate cluster count; the minimum is achieved at $k=7$.}
\label{tab:bic-scores}
\begin{tabular}{@{}cc@{}}
\toprule
\textbf{$k$} & \textbf{BIC} \\
\midrule
2 & 23{,}468.66 \\
3 & 23{,}234.48 \\
4 & 23{,}058.79 \\
5 & 22{,}877.57 \\
6 & 22{,}714.41 \\
7 & 22{,}604.61 \\
\bottomrule
\end{tabular}
\end{table}
\begin{figure}[t]
\centering
\includegraphics[width=0.8\linewidth]{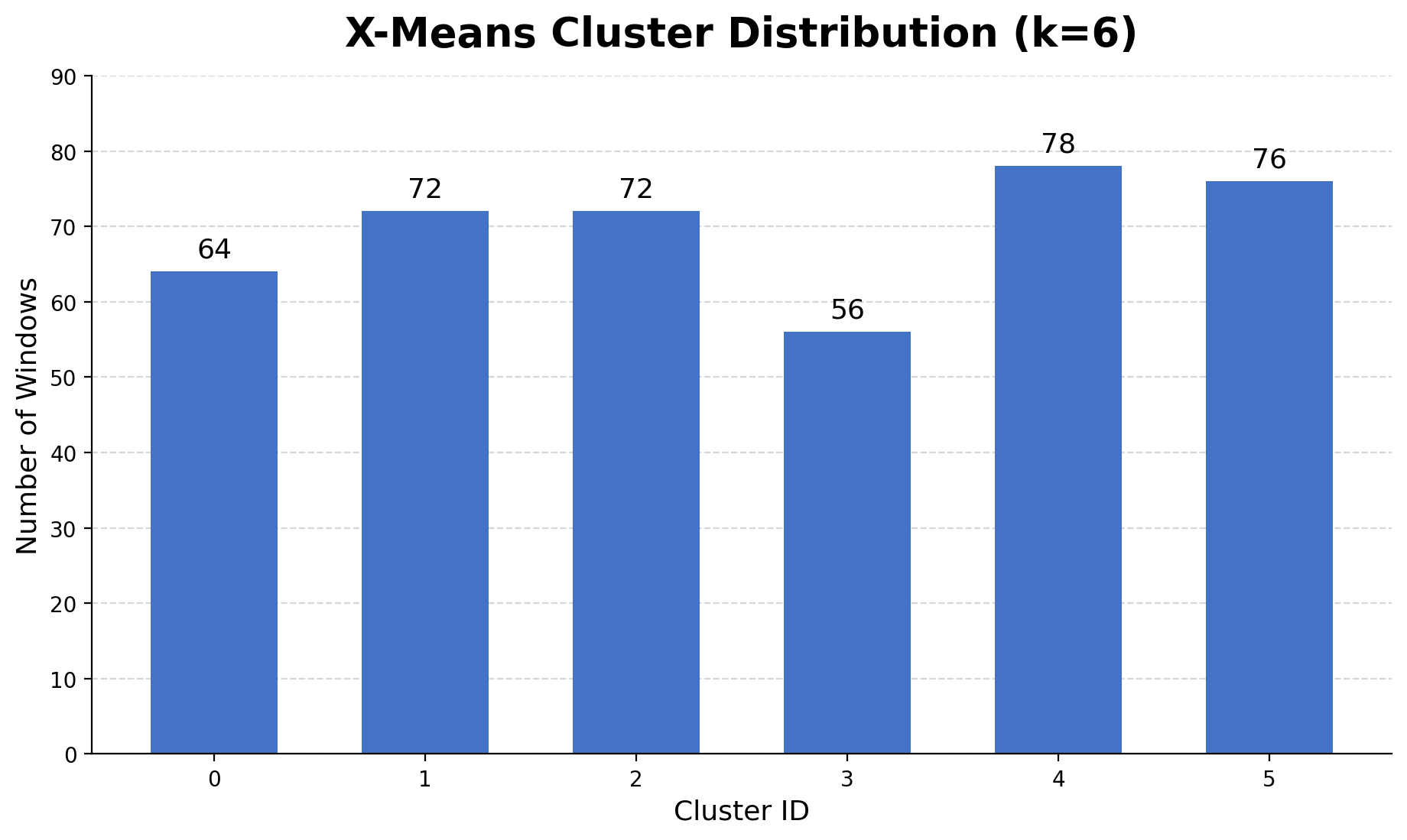}
\caption{Distribution of benign windows across six X-means clusters (56--78 windows per cluster), merged down from BIC's $k=7$.}
\label{fig:cluster-distribution}
\end{figure}

The number of distinct benign behavior modes in this reduced space is not known a priori, so it is learned with X-means: $k$-means is run for a range of candidate cluster counts, each scored with the Bayesian Information Criterion (BIC), which trades off explanatory power against model complexity, and the count with the lowest BIC is retained. The reference work found clustering-based one-class models to outperform Gaussian mixture models and one-class SVMs for this task~\cite{hamza2019detecting}, motivating our use of the same approach. The maximum candidate cluster count was capped at seven, matching the number of benign protocol categories generated by the traffic generator (Section~\ref{subsec:benign-traffic}); we verified this domain-informed cap experimentally by also running X-means without it. Uncapped, BIC is actually minimized at $k=16$ (BIC~$=22{,}346.51$, versus $22{,}374.61$ at $k=14$); however, two of the resulting sixteen clusters retain only one and three windows respectively and are merged into their nearest larger cluster under the same minimum-cluster-size rule used in the reported deployment, leaving 14 effective clusters after merging. On our 345-window training set, this uncapped configuration raised the false-positive rate from 2.15\% to 4.78\% due to over-clustering (this ablation is measured on the training set itself and is therefore not directly comparable to the 2.92\% held-out false-positive rate reported in Section~\ref{subsec:detection-performance}; both use the same clustering procedure, but different data splits). In the reported deployment, BIC is minimized at $k=7$ (Table~\ref{tab:bic-scores}); one of these seven raw clusters retains only four windows and is merged into its nearest larger cluster under the same minimum-cluster-size rule, yielding six stable clusters with 56--78 windows each (Figure~\ref{fig:cluster-distribution}); any cluster retaining fewer than ten windows after selection is merged into its nearest larger cluster by Manhattan distance, to avoid unstable boundary estimates from sparsely populated clusters.

For each of the six clusters, a boundary distance is computed as the 99th percentile of the Manhattan distance from every training and calibration window to its nearest cluster center -- a stricter threshold than the 97.5th percentile used in the reference work~\cite{hamza2019detecting}, adopted here to further reduce the false-positive rate. At inference time, a window's distance to its nearest cluster center is divided by that cluster's boundary distance; a resulting ratio above one places the window outside the cluster's learned normal region.

\subsection{Score Fusion}
\label{subsec:score-fusion}
\begin{figure}[t]
\centering
\includegraphics[width=0.95\linewidth]{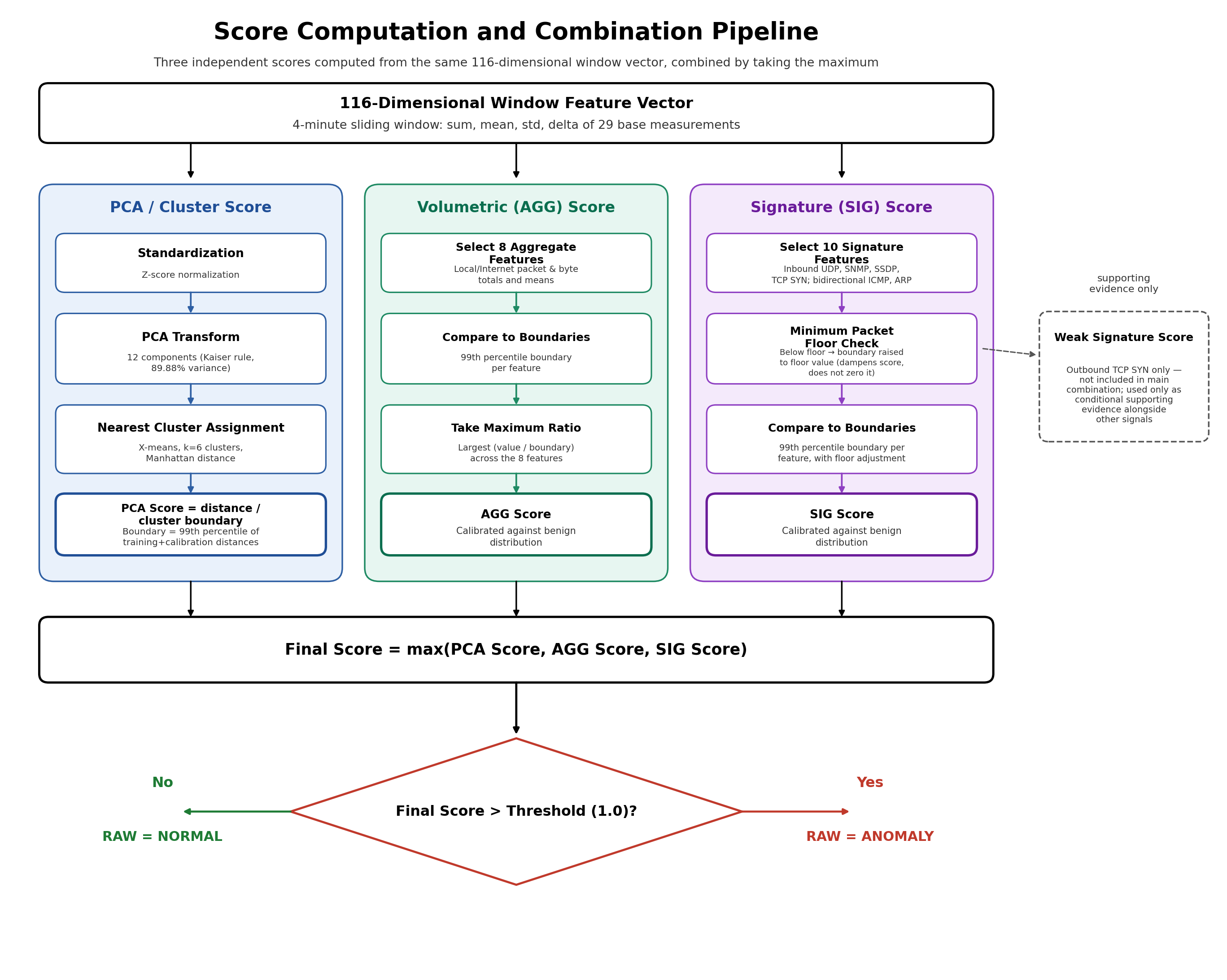}
\caption{Score computation and fusion pipeline: the cluster-based (PCA) score, the volumetric score, and the signature score are each normalized against a benign-calibrated boundary, and the final score is their maximum.}
\label{fig:score-pipeline}
\end{figure}

A single cluster-based score is not sufficient on its own: a dense but narrowly scoped attack can be diluted when averaged across all 116 dimensions simultaneously. Two additional scores are computed directly from the standardized raw features, without passing through the PCA/clustering step, and are combined with the cluster-based score to form the window's final score (Figure~\ref{fig:score-pipeline}).

The \emph{volumetric score} is computed from aggregate inbound/outbound packet and byte features: a 99th-percentile boundary is estimated per feature from the benign windows, each feature's value in a new window is divided by its boundary, and the largest such ratio across all features is taken as the volumetric score, capturing attacks that manifest primarily as a spike in traffic volume rather than as a shift in the overall behavioral profile.

The \emph{signature score} captures protocol- and port-specific patterns that do not necessarily involve a volume spike. Signatures are split into strong and weak evidence. Strong signatures cover inbound UDP traffic, bidirectional SNMP and SSDP port traffic, inbound TCP SYN packets, and bidirectional ICMP and ARP traffic; inbound-only tracking is used for UDP because outbound UDP (DNS/NTP queries) is a normal part of benign IoT behavior and would otherwise generate excessive false positives, whereas SNMP, SSDP, ICMP, and ARP are tracked bidirectionally because reflection and spoofing attacks can produce abnormal traffic in either direction. Outbound TCP SYN activity is tracked separately as weak evidence, since it also occurs during legitimate cloud-connectivity attempts and is only used to corroborate other signals rather than to trigger an alarm on its own. Only packet counts, not byte counts, feed the signature score, since volume information is already captured by the volumetric score. Signature boundaries are computed the same way as volumetric boundaries (99th percentile of benign windows), with one adjustment: for sparse protocols such as SNMP and ARP, a benign window may contain only a handful of packets, which would otherwise produce a boundary so low that the normalized score during an attack becomes disproportionate to the actual threat; an empirically determined minimum packet-count floor is applied to the boundary in these cases.

The final score for a window is the maximum of the cluster-based score, the volumetric score, and the strong-signature score; the weak-signature score is excluded from this maximum and is instead used conditionally by the live monitor (Section~\ref{subsec:live-monitoring}) as corroborating evidence. Because every component score is normalized so that the benign distribution's calibrated boundary corresponds to 1.0, a final score above 1.0 has a single, consistent interpretation regardless of which component triggered it: the window falls outside the region of behavior the model was trained to consider normal. The detection threshold is therefore not a separately tuned parameter but a direct consequence of this normalization, and calibration ensures that 99\% of benign data falls below it.

\subsection{Live Monitoring and Alarm State Machine}
\label{subsec:live-monitoring}
The trained model -- standardization parameters, PCA transform, cluster centers and boundaries, and volumetric/signature boundaries and calibration constants -- is serialized and loaded by a live monitoring component that applies the identical feature-computation logic used during training to each newly closed window, requiring no retraining to produce predictions. The live capture path reuses the same BPF filter and 5\,ms packet-deduplication logic as the offline pipeline (Section~\ref{subsec:data-collection}) for consistency, with a periodic cleanup routine that evicts stale packet- and flow-tracking table entries to bound memory growth during long-running deployment. A one-second control loop finalizes each closed window into its 29-dimensional row and appends it to a four-window rolling buffer; no score is produced until this buffer first fills, after which every newly closed window yields a score continuously.

A single anomalous window does not by itself raise an alarm. To suppress transient noise, a four-state alarm state machine -- normal, pre-alert, alert-onset, and sustained alert -- requires three consecutive anomalous windows before transitioning into the alert-onset state, which is the event that triggers the automated response pipeline (Section~\ref{sec:response-pipeline}). Returning to the normal state requires both five consecutive normal windows \emph{and} that at least 180 seconds have elapsed since the alert began; requiring both conditions jointly prevents short-lived returns to normal from causing the alarm to rapidly toggle open and closed.

\begin{figure}[t]
\centering
\includegraphics[width=0.95\linewidth]{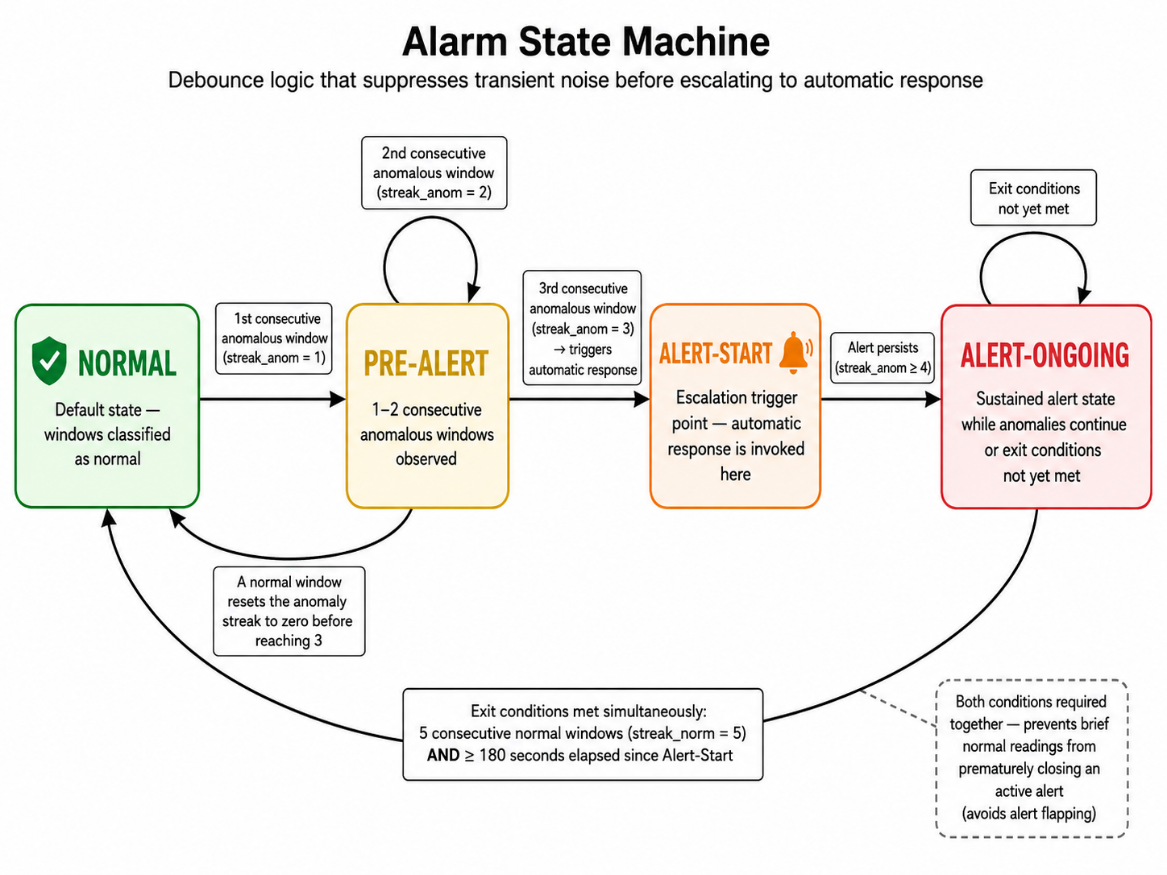}
\caption{Alarm state machine. Three consecutive anomalous windows raise an alert, which is only cleared after five consecutive normal windows and a minimum 180-second dwell time.}
\label{fig:alarm-state-machine}
\end{figure}

\section{Automated Response Pipeline}
\label{sec:response-pipeline}

\begin{figure*}[t!]
\centering
\includegraphics[width=0.85\textwidth]{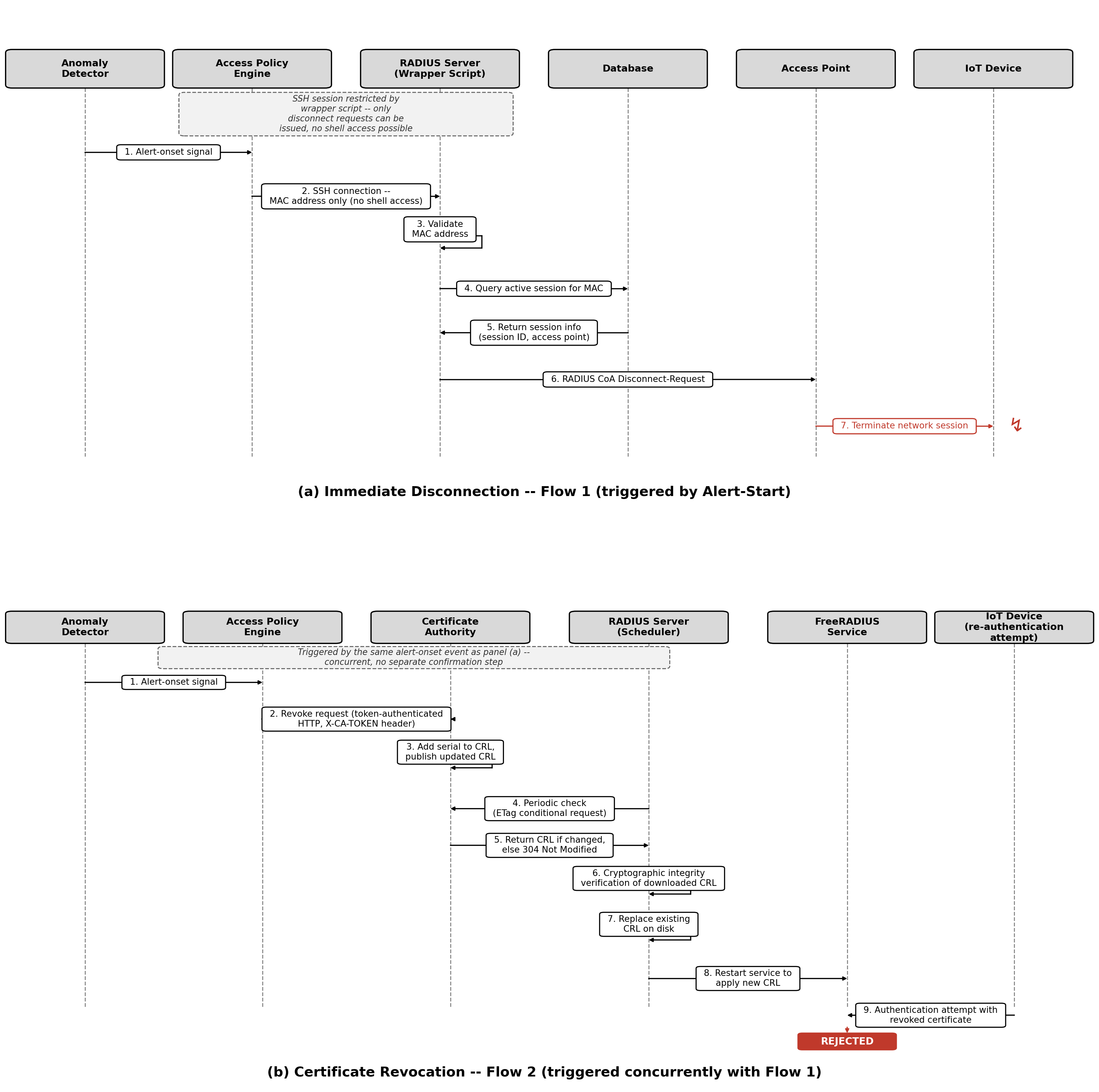}
\caption{Automated response flows, both triggered by the same alert-onset transition (Section~\ref{subsec:live-monitoring}), with no separate confirmation step between them. (a) Immediate disconnection: the access policy engine reaches the RADIUS server through a restricted SSH wrapper that accepts only a MAC address, looks up the active session, and issues a RADIUS CoA Disconnect-Request to the access point. (b) Certificate revocation: the access policy engine requests revocation via a token-authenticated endpoint on the certificate authority; the resulting, periodically fetched and integrity-checked CRL replaces the RADIUS server's revocation list, after which FreeRADIUS is restarted and future authentication attempts from the revoked device are rejected.}
\label{fig:response-flows}
\end{figure*}

This section describes the concrete, tested instantiation of the anomaly-driven response introduced in Section~\ref{subsec:response-channel}: the two-stage response that fires when the alarm state machine (Section~\ref{subsec:live-monitoring}) transitions into the alert-onset state, resolving policy rule P-03 (Table~\ref{tab:network-policies}). Both stages are issued automatically by the same response script on the same alert-onset transition, with no operator approval in the loop; the distinction between them is not that one requires confirmation and the other does not, but that they act on different time horizons -- an immediate, reversible network eviction, and a certificate revocation whose effect is durable across reauthentication attempts.

\subsection{Device Enrollment}
Every IoT device in the lab network is enrolled under IEEE 802.1X with EAP-TLS, each holding a unique client certificate issued by a lab certificate authority (Section~\ref{subsec:network-level}). When a device joins the network, the access point forwards the authentication request to a FreeRADIUS server, which validates the certificate and checks it against the current revocation list before granting or denying access. A successful authentication results in a session record -- MAC address, session identifier, and access point -- being stored in a database; this record is what the response pipeline looks up when it needs to act on a specific device.

\subsection{Instant Disconnection}
\label{subsec:instant-disconnect}
This is the concrete implementation of the response channel introduced in Section~\ref{subsec:response-channel}. When the alarm state machine signals alert-onset, the responding device's MAC address is resolved automatically from system records, and a response script is invoked with that address as its only input. Depending on configuration, this script runs either synchronously, blocking the monitoring loop until it completes, or asynchronously in the background so that monitoring continues uninterrupted; the choice is a deployment-time tradeoff between simplicity and monitoring continuity rather than one that affects the resulting network action.

The response script connects to the RADIUS server over a pre-configured SSH channel and passes only the target MAC address. On the server side, this SSH session is constrained by a wrapper script such that the only operation the detection host can invoke remotely is a disconnect request -- no interactive shell access is possible through this channel. The wrapper validates the MAC address, looks up the corresponding active session in the database, and issues a RADIUS Change-of-Authorization Disconnect-Request~\cite{chiba2008rfc5176} to the access point that is currently serving the device, together with the session information needed to identify the exact session to terminate. On receipt, the access point immediately tears down the device's network session (Figure~\ref{fig:response-flows}(a)). This action is intentionally reversible: a device that reauthenticates after being disconnected is admitted again under the normal 802.1X flow, unless it has also been placed on the certificate revocation list described next.

\subsection{Certificate Revocation and Permanent Exclusion}
\label{subsec:cert-revocation}
Instant disconnection removes a device from the network but does not, by itself, prevent it from reauthenticating a moment later. The response script therefore also revokes the device's certificate, issued automatically alongside the disconnect action rather than gated on any further confirmation, by calling a token-authenticated HTTP endpoint on the certificate authority (validated via an \texttt{X-CA-TOKEN} header) with the device's certificate serial number. A certificate revocation list (CRL) is maintained by the certificate authority and republished when changed; a timer process on the RADIUS server periodically fetches the current CRL, using an ETag-based conditional request so that an unchanged list is not needlessly re-downloaded.

When a new CRL is retrieved, it is first subjected to a cryptographic integrity check before replacing the server's current list, and FreeRADIUS is restarted to bring the updated list into effect (Figure~\ref{fig:response-flows}(b)) -- a full restart is required because FreeRADIUS loads the CRL into memory only at process startup, and a configuration reload (\texttt{systemctl reload}) was verified in lab testing not to refresh this in-memory copy, leaving a revoked certificate acceptable until a restart actually occurs. From that point on, any authentication attempt presenting the revoked certificate is rejected at the EAP-TLS handshake, so the device is excluded from the network regardless of how many times it attempts to reconnect, until the certificate is explicitly reinstated by an operator.

\subsection{Relationship to the Policy Model}
The two stages above are what rule P-03's ``restrict or terminate'' action (Table~\ref{tab:network-policies}) resolves to in practice: an anomaly-driven alert-onset transition triggers both the disconnect step of Section~\ref{subsec:instant-disconnect} and certificate revocation (Section~\ref{subsec:cert-revocation}) together, automatically and without a separate confirmation step.

\section{Experimental Setup}
\label{sec:experimental-setup}

\subsection{Testbed}
\label{subsec:testbed}
\begin{figure*}[t]
\centering
\begin{tikzpicture}[
  box/.style={draw, rounded corners, align=center, minimum height=0.8cm, text width=2.6cm, font=\scriptsize},
  arr/.style={-{Latex[length=1.8mm]}, thick}
]
\node[box] (internet) at (0,0) {Internet};
\node[box] (fw) at (3.1,0) {pfSense\\(firewall / VPN GW)};
\node[box] (sw) at (6.2,0) {Switch};
\node[box, text width=3.2cm] (esxi) at (9.6,1.1) {ESXi host: Kali (attacker), Wazuh, FreeRADIUS, CA, Android/Win11/Ubuntu VMs};
\node[box, fill=gray!15] (dut) at (9.4,-0.6) {Device under test: Raspberry Pi (PIR motion)};
\node[box] (ap) at (12.6,-0.6) {AP (other RPi nodes, industrial IoT devices)};

\draw[arr] (internet) -- (fw);
\draw[arr] (fw) -- (sw);
\draw[arr] (sw) -- (esxi);
\draw[arr] (sw) -- (dut);
\draw[arr] (sw) .. controls +(2.0,0) and +(-2.5,0.8) .. (ap);
\end{tikzpicture}
\caption{Testbed topology. Attacks are launched from the Kali VM toward the device under test; benign traffic is generated on the device itself (Section~\ref{subsec:benign-traffic}).}
\label{fig:testbed}
\end{figure*}
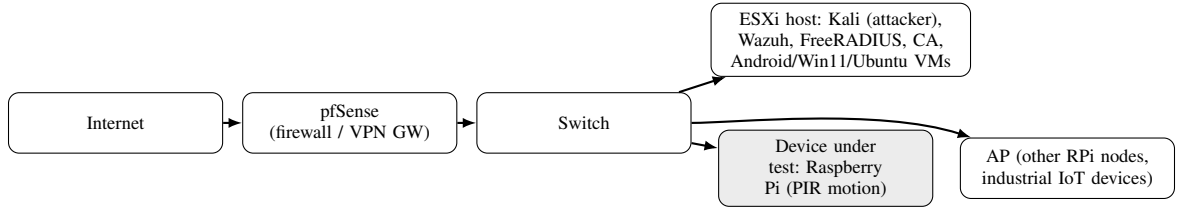

The testbed (Figure~\ref{fig:testbed}) is a lab network behind a pfSense firewall/VPN gateway, which mediates both site-to-site VPN and WAN/Internet connectivity. Behind the firewall, a single 1U server running VMware ESXi hosts the virtual machines used for orchestration and analysis: an Android VM and a Windows~11 VM used for general client emulation, an Ubuntu VM, a Wazuh instance used for centralized logging/SIEM, a FreeRADIUS server VM and a certificate authority (CA) server VM that together implement the network access-control and certificate-revocation mechanisms of Sections~\ref{sec:architecture} and~\ref{sec:response-pipeline}, and a Kali Linux VM used as the attack source. A managed switch connects the ESXi host to a set of Raspberry Pi nodes, each carrying a different sensor (humidity/temperature/color, passive-infrared motion detection, or ultrasonic distance sensing), and to a wireless access point. The device under test in this paper is a single Raspberry Pi node equipped with a passive-infrared motion sensor; the remaining Raspberry Pi nodes and a set of newly added industrial IoT devices reachable through the access point (a Wi-Fi printer, a digital photo frame, an air-quality sensor, and a smart smoke sensor) are part of the same physical testbed but were not included in the evaluation reported here (Section~\ref{sec:limitations}).

\subsection{Data Collection Protocol}
Following the phased protocol described in Section~\ref{subsec:data-collection}, the device under test was first observed for an extended benign-only period to build the training and calibration sets, after which eight attack types -- ARP spoofing, TCP SYN flooding, UDP flooding, Ping-of-Death, SNMP reflection, SSDP reflection, TCP SYN reflection, and Smurf (Section~\ref{subsec:attack-scenarios}) -- were each launched from the Kali VM at three intensities (1, 10, and 100 packets/second), for a total of 24 attack scenarios, each lasting 120 seconds and each separated by a transition phase. The resulting 116-dimensional feature vectors (Section~\ref{subsec:features}) were split as described in Section~\ref{subsec:training}: a chronologically ordered 70\%/15\%/15\% split of the safe benign windows into training, calibration, and held-out evaluation sets, yielding 345 training windows for the reported model.

\subsection{Attack Scenarios}
\label{subsec:attack-scenarios}
The eight attack types fall into two categories. \emph{Direct attacks} target the device under test directly: ARP spoofing floods the local network with ARP requests impersonating the device; TCP SYN flooding sends a stream of SYN packets to the device's TCP port without completing the handshake; UDP flooding directs a high rate of UDP packets at the device; and Ping-of-Death sends oversized ICMP echo requests. \emph{Reflection attacks} spoof the device's IP address to elicit disproportionately large responses from third parties toward it or, when the device's address is the spoofed source, toward other machines on its behalf: SNMP reflection sends GetBulk queries to the device's SNMP port with a spoofed source address; SSDP reflection sends M-SEARCH requests to the device's UPnP port with a spoofed source address; TCP SYN reflection spoofs the device's address toward a local reflector, causing SYN-ACK responses (and consequent RST replies) to arrive at the device; and the Smurf attack spoofs the device's address as the source of ICMP echo requests broadcast to the local subnet, causing every responding host to flood the device with replies.

\subsection{Evaluation Protocol}
Two complementary evaluations were performed. First, a continuous, purely benign session of approximately 137 minutes was recorded through the live monitoring pipeline (Section~\ref{subsec:live-monitoring}), yielding 137 windows with no attack activity, used to measure the false-positive rate under realistic, extended benign operation. Second, all 24 attack scenarios were replayed through the same live pipeline; because each 120-second attack is observed through a four-minute rolling window, its effect spans five consecutive scored windows (window size plus attack duration, minus one for the overlap), yielding 118 attack windows in total. The combined set of 137 benign and 118 attack windows is the basis for the confusion matrix, per-attack breakdown, and ROC analysis reported in Section~\ref{sec:results}.

\section{Results}
\label{sec:results}

\subsection{Detection Performance}
\label{subsec:detection-performance}
All 24 attack scenarios were successfully detected at the scenario level (24/24). At the finer-grained window level, 3 of the 118 attack windows were missed; all three follow the same pattern and occur at the first window in which a given attack enters the four-minute rolling buffer, where the window contains only one of the two feature ``buckets'' contributed by a still-ongoing attack and the accumulated signal has not yet crossed the threshold. We interpret this as an expected transient of the windowing mechanism rather than a missed attack, since every attack is subsequently flagged in each of its remaining four windows.

\begin{figure}[t]
\centering
\includegraphics[width=0.85\linewidth]{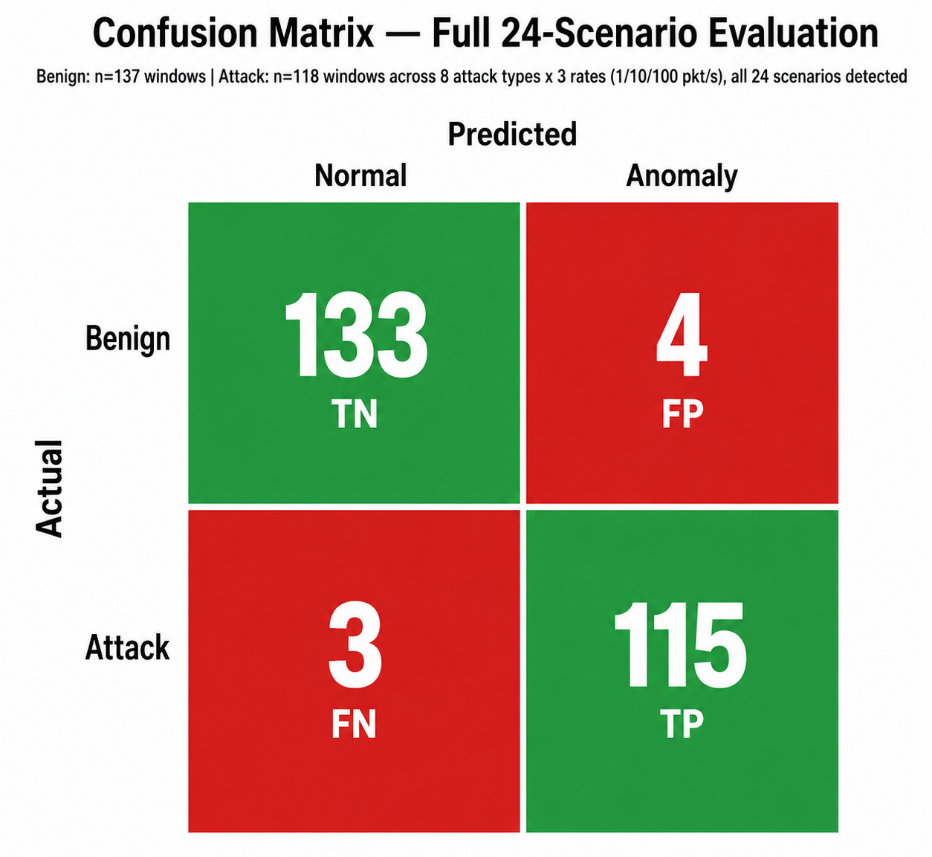}
\caption{Confusion matrix over the combined set of 137 benign and 118 attack windows.}
\label{fig:confusion-matrix}
\end{figure}

\begin{figure}[t]
\centering
\includegraphics[width=0.9\linewidth]{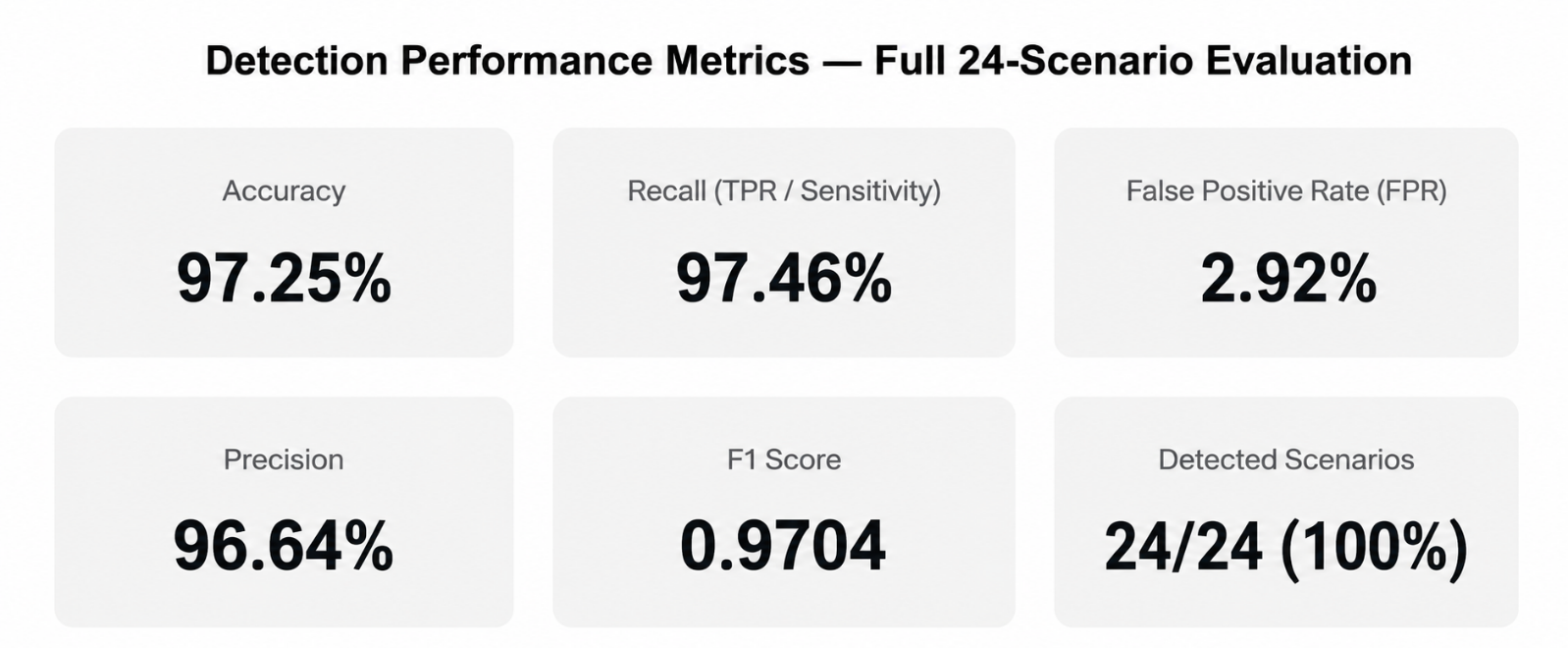}
\caption{Overall performance metrics (accuracy, precision, recall, F1) on the combined benign/attack window set.}
\label{fig:performance-metrics}
\end{figure}

\begin{figure}[t]
\centering
\includegraphics[width=0.95\linewidth]{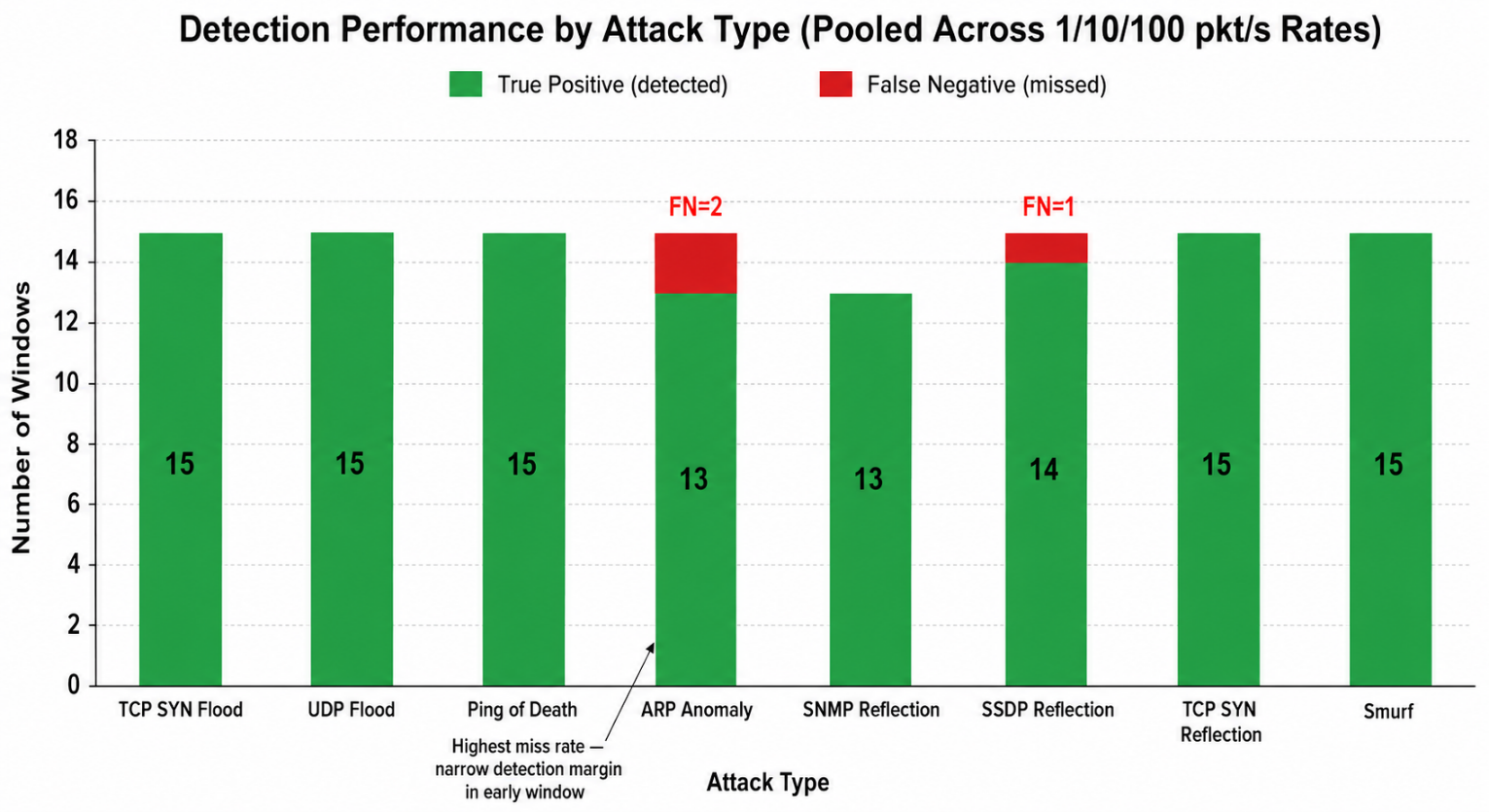}
\caption{Detection performance broken down by attack type.}
\label{fig:detection-by-attack}
\end{figure}

Figure~\ref{fig:confusion-matrix} shows the resulting confusion matrix and Figure~\ref{fig:performance-metrics} the associated performance metrics. Figure~\ref{fig:detection-by-attack} breaks detection down by attack type; the lowest recall is observed for the ARP-based attack (13/15 windows correctly flagged), which we attribute to ARP traffic's inherently low and variable benign baseline -- the minimum-packet-count floor applied to sparse-protocol signature boundaries (Section~\ref{subsec:score-fusion}) slightly delays the signal in the earliest windows of this particular attack.

\begin{figure}[t]
\centering
\includegraphics[width=0.85\linewidth]{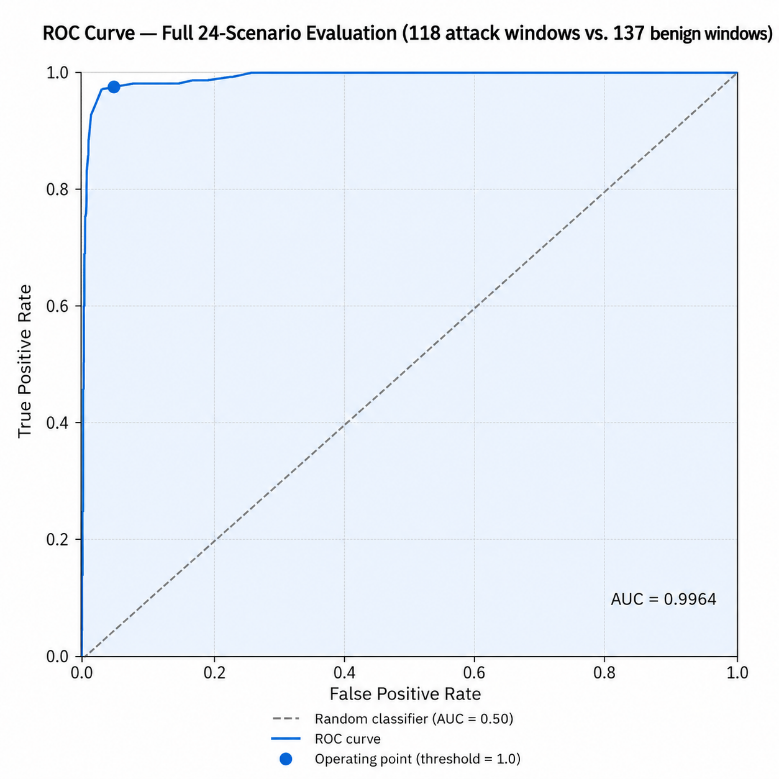}
\caption{ROC curve over the combined 118 attack / 137 benign windows (AUC~$=0.9964$). The operating point corresponding to the deployed threshold (score~$=1.0$) is marked at FPR~$=2.92\%$, TPR~$=97.46\%$.}
\label{fig:roc-curve}
\end{figure}

Figure~\ref{fig:roc-curve} shows the ROC curve computed over the same combined window set, with an area under the curve of 0.9964, indicating that the three fused scores separate benign from attack windows with high fidelity independent of any specific threshold choice. The deployed threshold of 1.0 -- which, as noted in Section~\ref{subsec:score-fusion}, follows directly from the calibration procedure rather than being separately tuned -- corresponds to an operating point of 2.92\% false-positive rate and 97.46\% true-positive rate, close to the top-left corner of the curve and consistent with a deliberately conservative balance that favors a low false-alarm rate given that a positive detection triggers an automated network action (Section~\ref{sec:response-pipeline}).

\subsection{Automated Response Latency}
\label{subsec:response-latency}
To quantify the response pipeline of Section~\ref{sec:response-pipeline} independently of the detection evaluation above, we measured end-to-end response latency on the same testbed using four attack scenarios -- TCP SYN flood, UDP flood, ARP spoofing, and SNMP reflection, each run at the lowest intensity (1 packet/second) -- against the trained model and the real disconnect-and-revoke response script. Two independent, millisecond-precision timestamp sources were used: the alarm state machine's own alert-onset transition, and timestamps logged by the response script at each of its steps.

\begin{table}[t]
\centering
\caption{Automated response latency across four attack scenarios (1 pkt/s; $n=4$).}
\label{tab:response-latency}
\small
\begin{tabular}{@{}p{0.55\linewidth}p{0.18\linewidth}p{0.18\linewidth}@{}}
\toprule
\textbf{Interval} & \textbf{Mean} & \textbf{Range} \\
\midrule
Alert-onset $\rightarrow$ network eviction (CoA disconnect) & 335.8\,ms & 319--372\,ms \\
Certificate-revocation HTTP round-trip & 111.5\,ms & 110--112\,ms \\
Total response script execution & 466.3\,ms & 449--504\,ms \\
\bottomrule
\end{tabular}
\end{table}

Table~\ref{tab:response-latency} reports the results. Across all four scenarios, the interval from alert-onset to the device's actual removal from the network -- measured as completion of the RADIUS CoA Disconnect-Request exchange -- averaged 335.8\,ms (range 319--372\,ms), with no evident dependence on attack type. Because RFC 5176 requires the access point to send its Disconnect-ACK only after the corresponding session has actually been torn down, this interval reflects the time to physically evict the device, not merely the time to transmit a message. The certificate-revocation HTTP round-trip to the certificate authority averaged 111.5\,ms (range 110--112\,ms) and varied little across attack types, consistent with a stable, low-noise request path. The two actions are issued concurrently by the response script, whose total execution time averaged 466.3\,ms (range 449--504\,ms) -- negligible relative to the 180-second alarm dwell time (Section~\ref{subsec:live-monitoring}) that already gates the response on sustained anomalous behavior. In all four trials, a subsequent reconnection attempt using the revoked certificate was rejected, independently confirmed via the RADIUS server's own access log.

\section{Discussion: Comparison with the Reference Architecture}
\label{sec:discussion}

Table-driven and prose comparisons throughout this paper have already contrasted individual design choices with those of Hamza et al.~\cite{hamza2019detecting}; this section draws the comparison together (Table~\ref{tab:reference-comparison}) and discusses what changes when a MUD/SDN-derived multi-model pipeline is replaced with the unified, passively captured design used here.

\begin{table*}[t]
\centering
\caption{Summary comparison between the reference architecture~\cite{hamza2019detecting} and the design used in this paper.}
\label{tab:reference-comparison}
\small
\begin{tabular}{@{}p{0.16\linewidth}p{0.40\linewidth}p{0.40\linewidth}@{}}
\toprule
\textbf{Parameter} & \textbf{Reference architecture~\cite{hamza2019detecting}} & \textbf{This work} \\
\midrule
Infrastructure & MUD profile + SDN (OpenFlow flow tables) & No MUD/SDN; passive packet capture \\
Model architecture & Two-stage (channel + flow workers), multi-model & Unified design over a single training set \\
Decision logic & Joint triggering of Stage-1 and Stage-2 workers within the same window & Maximum of cluster-based, volumetric, and signature scores \\
Flow decomposition & Separate flow tracking per MUD-derived rule & Single unified feature vector by protocol/direction/port \\
False-positive reduction & Dual-confirmation (two-stage) triggering mechanism & Alarm state machine (3 consecutive anomalies, $\geq$180\,s dwell) \\
Window size & Tested 1--8 min; 4 min selected & 4 min (adopted per reference finding) \\
Feature statistics & Sum, mean, standard deviation & Sum, mean, standard deviation, delta \\
Dimensionality reduction & PCA, Kaiser rule (2--18 components per worker) & PCA, Kaiser rule (single model, 12 components) \\
Clustering & X-means, 2--53 clusters per worker & X-means, $k=7$ via BIC $\rightarrow$ merged to 6 \\
Boundary percentile & 97.5\% & 99.0\% \\
Distance metric & Manhattan & Manhattan (same) \\
Training data scale & $\sim$15{,}000 min/device ($\approx$250\,h); required by the multi-model structure & 345 windows ($\approx$5.8\,h); single-model architecture reaches reliable results at this scale \\
Attack types & 8 types (4 direct, 4 reflection) & 8 types (4 direct, 4 reflection) \\
Attack rates & 1, 10, 100 packets/second & 1, 10, 100 packets/second \\
Reported performance (aggregate) & Accuracy 94.9\% / TPR 89.7\% / FPR 5.1\% & Accuracy 97.25\% / TPR 97.46\% / FPR 2.92\% \\
\bottomrule
\end{tabular}
\end{table*}

In the reference architecture, Stage-2 workers are bound one-to-one to OpenFlow rules derived from a device's MUD profile, each with its own independently trained model over its own packet/byte counters, while Stage-1 workers separately model the device's local and Internet-facing channels; an alert requires a Stage-1 worker and a Stage-2 worker to agree within the same time window, and a Markov-chain component additionally checks whether the sequence of cluster transitions itself looks unusual. This design distributes the one-class learning problem across dozens of narrowly scoped models, each benefiting from the fine-grained visibility an OpenFlow switch provides into per-flow behavior. Because our design has no SDN flow table to draw such per-flow telemetry from (Section~\ref{subsec:why-not-sdn}), traffic is instead classified only by protocol, direction, and service port from a passive capture, which rules out training one independent model per MUD-derived flow rule. We therefore apply PCA and X-means clustering once, over the full 116-dimensional feature vector, producing a single cluster-based score for the device's overall behavior, complemented by two additional scores -- volumetric and signature-based -- computed directly from raw feature percentiles without passing through the clustering step. In effect, the role played by dozens of independently trained Stage-1/Stage-2 workers in the reference design is redistributed across three parallel, mutually substitutable scores, any one of which crossing its calibrated boundary raises the final score past the detection threshold.

A direct consequence of this redistribution is a substantial reduction in the training data the model needs. The reference architecture trains its ensemble of per-flow models on roughly 15{,}000 minutes of benign traffic per device; the deployment reported here reaches comparable detection performance (Section~\ref{subsec:detection-performance}) from 345 training windows -- a difference of roughly 43-fold ($15{,}000/345 \approx 43.5$). This is consistent with the intuition that a single, unified model with a modest number of parameters requires less data to fit reliably than an ensemble of independently trained, per-flow models, though we note this comparison is illustrative rather than a controlled, head-to-head experiment on the same dataset (Section~\ref{sec:limitations}). Despite this reduction in training data, the aggregate detection performance reported for our deployment (97.25\% accuracy, 97.46\% TPR, 2.92\% FPR; Section~\ref{subsec:detection-performance}) is at least as strong as the aggregate figures Hamza et al.\ report across their own device set (94.9\% accuracy, 89.7\% TPR, 5.1\% FPR)~\cite{hamza2019detecting}, though, as noted below, the two figures come from different testbeds and device populations and should not be read as a controlled comparison.

The reference design's dual Stage-1/Stage-2 confirmation requirement is, functionally, a mechanism for suppressing false positives by requiring agreement across two independent views of the traffic before raising an alert. Lacking two independent per-flow views, we instead rely on the alarm state machine's temporal hysteresis (Section~\ref{subsec:live-monitoring}): requiring three consecutive anomalous windows before an alert is raised, and both five consecutive normal windows and a minimum dwell time before it clears, plays an analogous role in suppressing transient, single-window false alarms, at the cost of adding a small, fixed detection delay rather than requiring cross-model agreement.

\subsection{Relationship to Commercial ITDR}
Section~\ref{subsec:itdr} noted that commercial WiFi identity threat detection and response (ITDR) products already drive RADIUS CoA and certificate revocation from an anomaly signal, using the same enforcement primitives as this paper. This is worth dwelling on, because it changes what the architectural part of our contribution should be read as. Given that industry has independently converged on RADIUS CoA and certificate revocation as practical, deployable enforcement actions, the interesting open question is no longer whether an anomaly signal can be turned into a network action this way -- it evidently can, at commercial scale -- but which anomaly signals feed that action. Identity-behavior ITDR and our network-traffic-behavior detector are complementary rather than competing: a device could be flagged by one and missed entirely by the other, since an ITDR baseline of login times and locations has no visibility into what a device does with its network access once authenticated, and our detector has no visibility into who or what is authenticating. A deployment combining both -- identity-behavior anomalies feeding one policy rule, traffic-behavior anomalies feeding another, both actuated through the same RADIUS-based access policy engine (Section~\ref{subsec:policy-model}) -- would plausibly catch a broader range of compromise than either alone; evaluating such a combination is beyond the scope of this paper but follows naturally from the contextual policy model already in place.

\section{Limitations and Threats to Validity}
\label{sec:limitations}

\textbf{Single-device evaluation.} All detection and response results in this paper (Sections~\ref{sec:experimental-setup}--\ref{sec:results}) were obtained from a single Raspberry Pi node running a passive-infrared motion sensor. The testbed (Section~\ref{subsec:testbed}) includes several other device types -- additional Raspberry Pi sensor nodes and newly added industrial IoT devices (a Wi-Fi printer, a digital photo frame, an air-quality sensor, and a smart smoke sensor) -- but these were not part of the present evaluation. We therefore make no claim that the detection performance reported in Section~\ref{subsec:detection-performance} generalizes across device types with different traffic profiles; extending the evaluation to this heterogeneous device set is the most immediate item of future work.

\textbf{Training-set size.} The reported model was trained on 345 benign windows, calibrated on a further 15\% holdout split, and evaluated on 137 benign and 118 attack windows drawn from a single extended data-collection session. While Section~\ref{subsec:detection-performance} shows this is sufficient to reach a high AUC on this device, the false-positive rate and detection latency could behave differently over longer deployments, seasonal variation in benign traffic, or network conditions not represented in this single session.

\textbf{Scope of the access policy engine.} The access policy engine is designed to be extensible to other access types under the same contextual rule model (Section~\ref{subsec:policy-model}), but this paper evaluates only the network access-control mechanism described in Section~\ref{sec:architecture}. We make no claims about the design or performance of other access types the engine may host.

\textbf{Attack diversity and adaptive adversaries.} The eight attack types evaluated (Section~\ref{subsec:attack-scenarios}) are representative volumetric and reflection attacks drawn from the reference work's own evaluation, run at fixed, non-adaptive intensities. An adversary aware of the detection thresholds and the 180-second alarm dwell time (Section~\ref{subsec:live-monitoring}) could, in principle, attempt to stay below the calibrated boundaries or exploit the hysteresis window; we did not evaluate the system's robustness against such adaptive, detection-aware adversaries.

\textbf{Comparison with the reference architecture is illustrative, not controlled.} The data-efficiency comparison in Section~\ref{sec:discussion} contrasts our training-set size with the figure reported by Hamza et al.~\cite{hamza2019detecting} for their own testbed and device set, rather than a head-to-head evaluation of both architectures on identical hardware and traffic. Differences in device behavior, network conditions, and attack implementations between the two studies mean this comparison should be read as indicative of the general effect of consolidating a multi-model pipeline into a single model, not as a controlled ablation.

\section{Conclusion}
\label{sec:conclusion}

We presented an access-control architecture that closes the loop between IoT network-behavior anomaly detection and concrete, automated access-control action, built entirely on standard protocols -- IEEE 802.1X, EAP-TLS, and RADIUS with dynamic authorization extensions -- rather than on a programmable data plane or manufacturer-published MUD profiles. A central access policy engine continuously consumes the anomaly detector's output and actuates a response against the device's network membership, and is designed to be extensible to other access types under the same contextual rule model. The architecture's contribution is deliberately in the enforcement path rather than in the detector: the one-class anomaly detector that triggers this response adapts an existing MUD/SDN-based design~\cite{hamza2019detecting} to a passive-capture setting, consolidating a multi-model pipeline into a single fused model that reached an AUC of 0.9964 and detected all 24 evaluated attack scenarios using roughly 43$\times$ less training data than the architecture it draws on. On a single testbed device, we demonstrated the full closed loop in practice: an anomaly signal reliably triggers an immediate RADIUS CoA-based disconnection, measured to complete in 335.8\,ms on average, with certificate revocation available as a durable, reauthentication-proof exclusion mechanism that completed a further 111.5\,ms later.

The evaluation reported here is intentionally scoped: it covers a single device type, a fixed set of attack scenarios, and the network access-control mechanism described in Section~\ref{sec:architecture}. Extending the evaluation across the heterogeneous device set already present in the testbed, evaluating robustness against detection-aware adaptive adversaries, and exploring how the same policy engine could be extended to other access types are the most direct next steps. More broadly, we hope this work makes the case that, for IoT security architectures, an evaluation of the response path deserves the same attention that detection accuracy conventionally receives.

\section*{Acknowledgment}
This work was supported in part by the Scientific and Technological Research Council of T\"urkiye (T\"UB\.ITAK) under Grant No.\ 3241032. The authors thank Murat Demirci from Y\"onsis A.\c{S}.\ for his technical support with the laboratory setup.

\section*{Data and Code Availability}
Code and data supporting this paper are available at: \url{https://github.com/Emir-Korkmaz/closing-the-loop-network-revocation}.

\bibliographystyle{IEEEtran}
\bibliography{references}

@misc{gartner2022itdr,
  author       = {{Gartner}},
  title        = {Gartner Identifies Top Security and Risk Management Trends for 2022},
  howpublished = {Gartner Newsroom Press Release},
  year         = {2022},
  month        = {March},
  url          = {https://www.gartner.com/en/newsroom/press-releases/2022-03-07-gartner-identifies-top-security-and-risk-management-trends-for-2022},
  note         = {Accessed 2026-07-05}
}

@misc{ironwifi2026itdr,
  author       = {{IronWiFi}},
  title        = {{WiFi} Identity Threat Detection {\&} Response ({ITDR})},
  howpublished = {Product documentation},
  year         = {2026},
  url          = {https://www.ironwifi.com/itdr/},
  note         = {Accessed 2026-07-05}
}

@inproceedings{hamza2019detecting,
  author    = {Hamza, Ayyoob and Gharakheili, Hassan Habibi and Benson, Theophilus A. and Sivaraman, Vijay},
  title     = {Detecting Volumetric Attacks on IoT Devices via {SDN}-Based Monitoring of {MUD} Activity},
  booktitle = {Proceedings of the 2019 ACM Symposium on SDN Research (SOSR '19)},
  year      = {2019},
  pages     = {36--48},
  address   = {San Jose, CA, USA},
  publisher = {ACM},
  doi       = {10.1145/3314148.3314352}
}

@inproceedings{hamza2018combining,
  author    = {Hamza, Ayyoob and Gharakheili, Hassan Habibi and Sivaraman, Vijay},
  title     = {Combining {MUD} Policies with {SDN} for {IoT} Intrusion Detection},
  booktitle = {Proceedings of the 2018 Workshop on IoT Security and Privacy (IoT S\&P '18)},
  year      = {2018},
  pages     = {1--7},
  address   = {Budapest, Hungary},
  publisher = {ACM},
  doi       = {10.1145/3229565.3229571}
}

@article{hamza2022verifying,
  author  = {Hamza, Ayyoob and Ranathunga, Dinesha and Gharakheili, Hassan Habibi and Benson, Theophilus A. and Roughan, Matthew and Sivaraman, Vijay},
  title   = {Verifying and Monitoring {IoT}s Network Behavior Using {MUD} Profiles},
  journal = {IEEE Transactions on Dependable and Secure Computing},
  volume  = {19},
  number  = {1},
  pages   = {1--18},
  year    = {2022},
  doi     = {10.1109/TDSC.2020.2997898}
}

@techreport{lear2019rfc8520,
  author      = {Lear, Eliot and Droms, Ralph and Romascanu, Dan},
  title       = {Manufacturer Usage Description Specification},
  institution = {IETF},
  number      = {RFC 8520},
  year        = {2019},
  doi         = {10.17487/RFC8520}
}

@techreport{chiba2008rfc5176,
  author      = {Chiba, Murtaza and Dommety, Gopal and Eklund, Mark and Mitton, David and Aboba, Bernard},
  title       = {Dynamic Authorization Extensions to Remote Authentication Dial In User Service ({RADIUS})},
  institution = {IETF},
  number      = {RFC 5176},
  year        = {2008},
  doi         = {10.17487/RFC5176}
}

@inproceedings{singh2019clearer,
  author    = {Singh, Simran and Atrey, Ashlesha and Sichitiu, Mihail L. and Viniotis, Yannis},
  title     = {Clearer than Mud: Extending Manufacturer Usage Description ({MUD}) for Securing {IoT} Systems},
  booktitle = {Internet of Things -- ICIOT 2019},
  series    = {Lecture Notes in Computer Science},
  volume    = {11519},
  year      = {2019},
  publisher = {Springer},
  address   = {Cham},
  doi       = {10.1007/978-3-030-23357-0_4}
}

@article{candal2020quarantining,
  author  = {Candal-Ventureira, David and Fondo-Ferreiro, Pablo and Gil-Casti{\~n}eira, Felipe and Gonz{\'a}lez-Casta{\~n}o, Francisco Javier},
  title   = {Quarantining Malicious {IoT} Devices in Intelligent Sliced Mobile Networks},
  journal = {Sensors},
  volume  = {20},
  number  = {18},
  pages   = {5054},
  year    = {2020},
  doi     = {10.3390/s20185054}
}

@article{meidan2018nbaiot,
  author  = {Meidan, Yair and Bohadana, Michael and Mathov, Yael and Mirsky, Yisroel and Shabtai, Asaf and Breitenbacher, Dominik and Elovici, Yuval},
  title   = {{N-BaIoT}---Network-Based Detection of {IoT} Botnet Attacks Using Deep Autoencoders},
  journal = {IEEE Pervasive Computing},
  volume  = {17},
  number  = {3},
  pages   = {12--22},
  year    = {2018},
  doi     = {10.1109/MPRV.2018.03367731}
}

@inproceedings{mirsky2018kitsune,
  author    = {Mirsky, Yisroel and Doitshman, Tomer and Elovici, Yuval and Shabtai, Asaf},
  title     = {Kitsune: An Ensemble of Autoencoders for Online Network Intrusion Detection},
  booktitle = {Proceedings of the Network and Distributed System Security Symposium (NDSS)},
  year      = {2018},
  doi       = {10.14722/ndss.2018.23204}
}

@inproceedings{antonakakis2017mirai,
  author    = {Antonakakis, Manos and April, Tim and Bailey, Michael and Bernhard, Matt and Bursztein, Elie and Cochran, Jaime and Durumeric, Zakir and Halderman, J. Alex and Invernizzi, Luca and Kallitsis, Michalis and Kumar, Deepak and Lever, Chaz and Ma, Zane and Mason, Joshua and Menscher, Damian and Seaman, Chad and Sullivan, Nick and Thomas, Kurt and Zhou, Yi},
  title     = {Understanding the {Mirai} Botnet},
  booktitle = {Proceedings of the 26th USENIX Security Symposium},
  year      = {2017},
  pages     = {1093--1110},
  address   = {Vancouver, BC, Canada},
  publisher = {USENIX Association}
}

@article{feraudo2020sok,
  author  = {Feraudo, Angelo and Yadav, Poonam and Mortier, Richard and Bellavista, Paolo and Crowcroft, Jon},
  title   = {{SoK}: Beyond {IoT} {MUD} Deployments -- Challenges and Future Directions},
  journal = {arXiv preprint arXiv:2004.08003},
  year    = {2020}
}

@misc{ieee8021x2020,
  title        = {{IEEE} Standard for Local and Metropolitan Area Networks -- Port-Based Network Access Control},
  howpublished = {IEEE Std 802.1X-2020},
  organization = {IEEE},
  year         = {2020},
  doi          = {10.1109/IEEESTD.2020.9018454}
}

\end{document}